\documentclass[%
 aip,
cp,  
 amsmath,amssymb,
 reprint,%
]{revtex4-2}

\usepackage{graphicx}
\usepackage{dcolumn}
\usepackage{bm}

\usepackage[utf8]{inputenc}
\usepackage[T1]{fontenc}
\usepackage{mathptmx}

\usepackage{subcaption}

\begin{document}

\title{Spin-Density Matrix Elements for Vector Meson Photoproduction at GlueX}

\author{Alexander Austregesilo} 
 \email[Corresponding author: ]{aaust@cmu.edu}
\affiliation{
  Canegie Mellon University, 5000 Forbes Ave, Pittsburgh, PA 15213, USA.
}

\author{the GlueX Collaboration}


\date{\today} 

\begin{abstract}
The GlueX experiment at Jefferson Lab aims to study the light meson spectrum with an emphasis on the search for hybrid mesons. To this end, a linearly-polarized $9\,$GeV photon beam impinges on a hydrogen target contained within a detector with near-complete neutral and charged particle coverage. In 2018, the experiment completed its first phase of data taking in its design configuration and the quantity of analyzed data already exceed that of previous experiments for polarized photoproduction in this energy regime by orders of magnitude. Polarization observables such as spin-density matrix elements provide valuable input for the theoretical description of the production  mechanism, which will be essential for the interpretation of possible exotic meson signals. We present results for the photoproduction of vector mesons, focusing on the unprecedented statistical precision of the spin-density matrix elements for the $\rho$(770) meson.
\end{abstract}

\maketitle

\section{Introduction}

\begin{figure}[h!]
\centerline{\includegraphics[width=.7\textwidth]{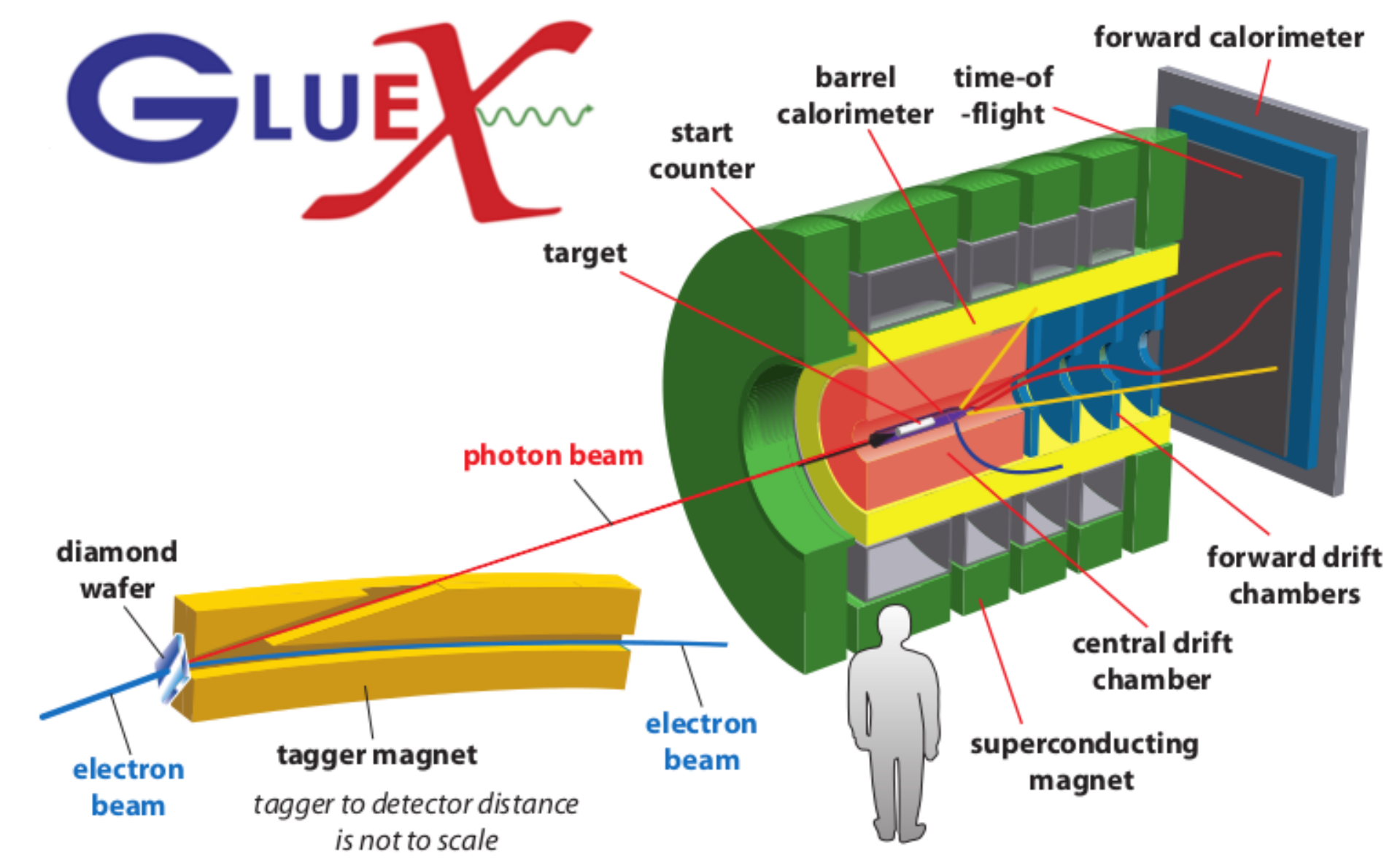}}
\caption{The GlueX detector.\label{fig:det}}
\end{figure}

The GlueX experiment at the Thomas Jefferson National Accelerator Facility is part of a global effort to study the spectrum of hadrons. A primary electron beam of up to 12\, GeV is used to produce a secondary photon beam which impinges on a liquid hydrogen target. The scattered electrons are used to tag the energy of the photon beam.  Assuming vector meson dominance, a wide variety of mesonic states are accessible. A high beam intensity provides a sufficiently large reaction rate to study rare processes. The GlueX detector was specifically designed to map the light quark meson spectrum up to masses of approximately 3\,GeV$/c^2$ with full acceptance for all decay modes. A superconducting solenoid magnet with a 2\,T field houses the target, a start counter~\cite{poo19}, central and forward drift chambers, and a barrel calorimeter~\cite{Bea18}. A forward calorimeter completes the forward photon acceptance and a time-of-flight counter provides particle identification capability (see~Fig.~\ref{fig:det}).

The first phase of the GlueX experiment was completed in  2018. About 20\% of the full data set was used to produce the results discussed here. From 2019 onwards, the upgraded detector will continue taking data with an even higher luminosity.

The unique feature of GlueX is the capability to use a polarized photon beam. Polarization of the photons is achieved by coherent Bremsstrahlung of the primary electron beam on a thin diamond radiator. With a collimator suppressing the incoherent Bremsstrahlung spectrum, a linear polarization of up to 40\% is achieved in the coherent peak at 9\,GeV. In order to cancel apparatus effects, the polarization plane is rotated in steps of 45$^\circ$ around the beam axis into four different orientations during physics data taking. The degree of polarization is measured using the effect of triplet production\cite{Dug17}. 

The photon beam polarization poses constraints on the quantum numbers of the produced meson systems. It may be used as a filter to enhance particular resonances or as an additional input in multidimensional amplitude analyses. To this end, the photoproduction mechanism has to be understood in great detail. Only very limited existing data from previous experiments are available at these energies. GlueX has already measured beam asymmetry observables for the production of the pseudoscalar mesons $\pi^0$ and $\eta$~\cite{Glx18} and is currently pursuing the analysis of $\eta'$ and $\pi^-$ photoproduction.

As an extension of this program, the following analysis aims to study the production mechanism for vector mesons. The vector mesons $\rho$(770), $\omega$(782) and $\phi$(1020) are produced abundantly at GlueX, and the angular distribution of their production and decay is fully described with the spin-density matrix elements (SDMEs) $\rho_{ij}^k$. The linear photon beam polarization provides access to nine linearly independent SDMEs~\cite{sch70}. The values and their dependence on the squared four-momentum transfer $-t$ constitute valuable constraints to the theoretical description of the process. Earlier measurements at SLAC~\cite{bal73} were based on only a few thousand events and generally reported a conservation of the helicity in the $s$-channel (SCHC), observing $\rho_{1-1}^1 = -\operatorname{Im} \rho_{1-1}^2 = 0.5$ in the helicity frame. All other SDMEs were roughly consistent with zero.
Even though the values were determined as a function of $-t$, the precision did not allow them to make any statements about the $t$ dependence. Nonetheless, they were recently combined with high energy data in Regge theory fits to produce precise theoretical predictions~\cite{mat18}, which will be compared to our measurements in this publication.

\section{Method}

We describe the analysis method for the production of the $\rho$(770) meson, which decays predominantly into $\pi^+\pi^-$. The analysis of $\omega$(782) and $\phi$(1020) follows the same general scheme.

\begin{figure}[ht]
    \begin{subfigure}{0.49\textwidth}
		\includegraphics[width=\textwidth]{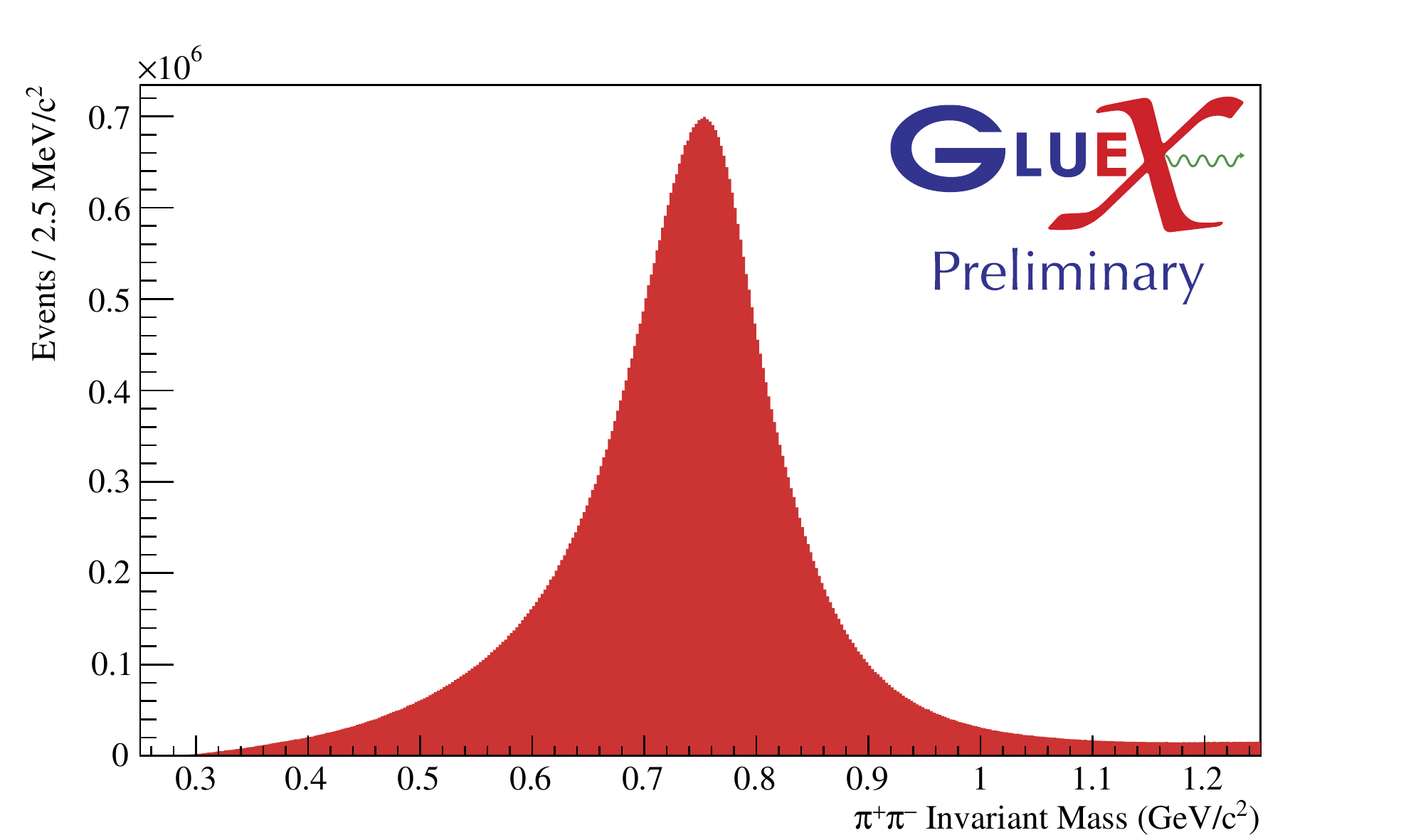}
		    \caption{$m_{\pi^+\pi^-}$}
	\end{subfigure}
	\begin{subfigure}{0.49\textwidth}
		\includegraphics[width=\textwidth]{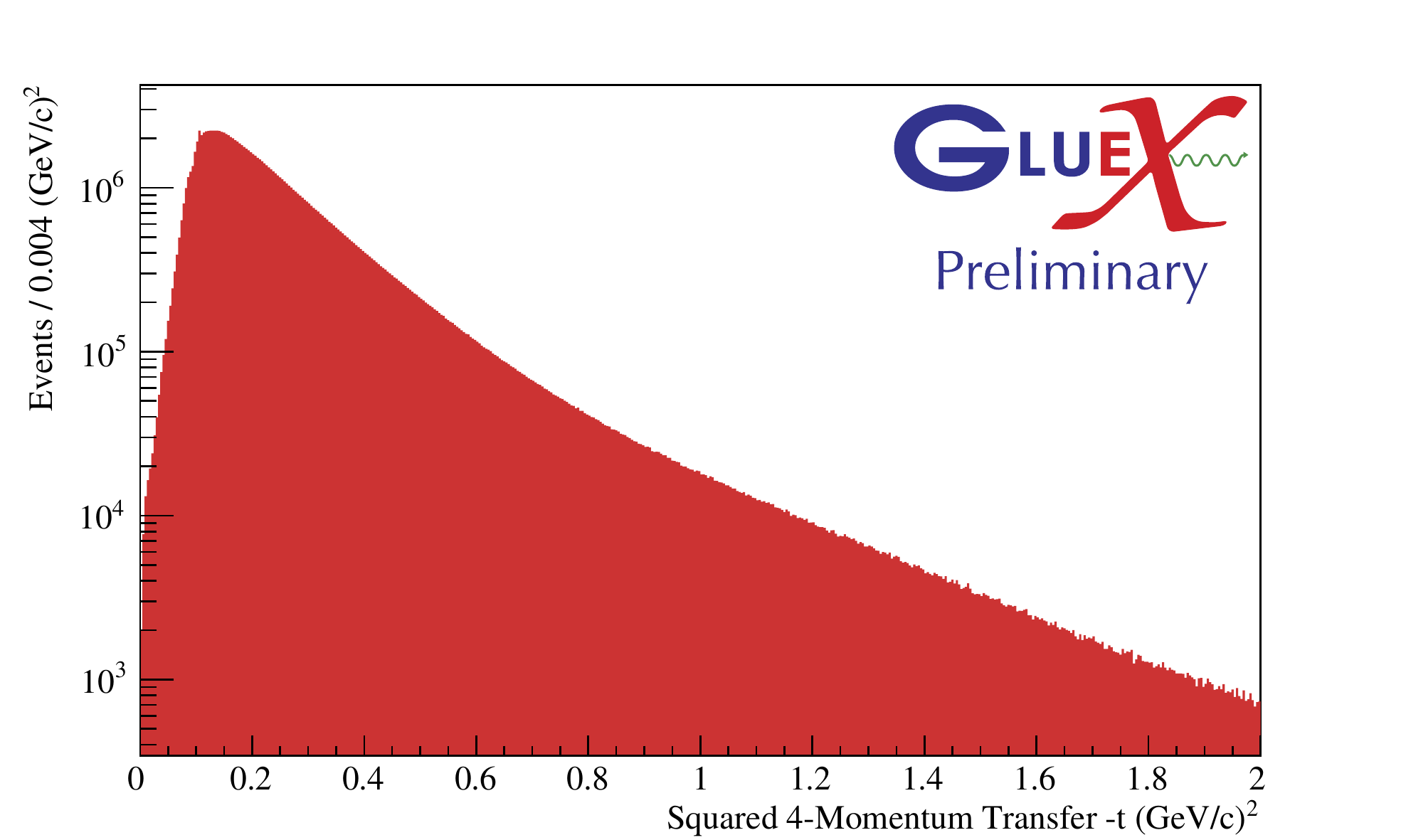}
		\caption{$-t$}
	\end{subfigure}
\caption{Kinematic distributions for the $\rho$(770) data sample.\label{fig:rho}}
\end{figure}

As a first step, we select exclusive events by completely reconstructing the final state ($\pi^+\pi^-p$) with all particle trajectories originating from the same vertex and requiring four-momentum conservation. The resulting invariant mass spectrum (see~Fig.~\ref{fig:rho}a) shows only a contribution from non-resonant background below the resonance peak on the order of a few percent. This background is neglected in the current stage of the analysis. Due to the requirement of a successfully reconstructed proton track, the distribution of the squared four-momentum transfer $-t$ shows a depletion at zero (see~Fig.~\ref{fig:rho}b). Since the acceptance is very low in this region, we discard all events with $-t$ below 0.05\,GeV$^{2}/c^2$. Even though we only analyze 20\% of our full data set, we have more than 10 million $\rho$(770) events available for each of the 4 orientations of the polarization.

\begin{figure}[ht]
\includegraphics[trim={145 412 200 125}, clip, width=.55\textwidth]{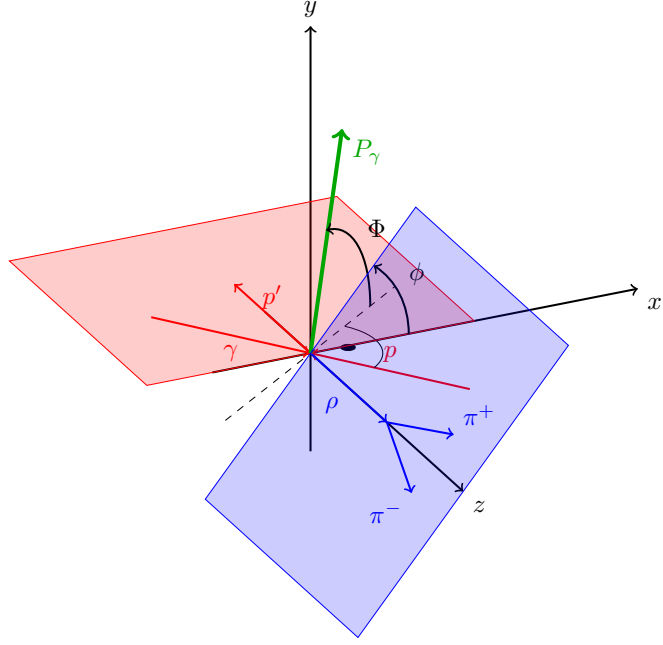}
\caption{\label{fig:coordinates} Definition of the angles relevant for the process of vector-meson photoproduction. The hadronic production plane is indicated in red, the $\rho$(770) decay plane in blue and the photon polarization vector in green.}
\end{figure}

The angular part of the measured intensity $W$ can be expressed as a function of three angles. The two decay angles $\vartheta$ and $\varphi$ are defined in the helicity frame of the vector meson (see~Fig.~\ref{fig:coordinates}). The direction of the photon polarization with respect to the hadronic production plane is defined by $\Phi$. Together with the independently measured degree of polarization $P_\gamma$, the angular part of the cross section for the production of a vector meson with a linearly polarized beam is expressed as follows:

\begin{eqnarray}\label{eq:sdme}
 W(\cos\vartheta, \varphi, \Phi) =&& W^0(\cos\vartheta, \varphi) - P_\gamma \cos(2\Phi) W^1(\cos\vartheta,\varphi) - P_\gamma\sin(2\Phi)W^2(\cos\vartheta, \varphi)\\
        W^0(\cos\vartheta, \varphi) =&& \frac{3}{4\pi} \left ( \frac{1}{2}(1-\rho^0_{00}) + \frac{1}{2}(3\rho^0_{00}-1)\cos^2\vartheta - \sqrt{2}\operatorname{Re}\rho^0_{10}\sin2\vartheta\cos\varphi-\rho^0_{1-1}\sin^2\vartheta\cos2\varphi \right )\nonumber\\
        W^1(\cos\vartheta, \varphi) =&& \frac{3}{4\pi} \left ( \rho^1_{11}\sin^2\vartheta + \rho^1_{00}\cos^2\vartheta - \sqrt{2}\operatorname{Re}\rho^1_{10}\sin2\vartheta\cos\varphi - \rho^1_{1-1}\sin^2\vartheta\cos2\varphi \right )\nonumber\\
        W^2(\cos\vartheta, \varphi) =&& \frac{3}{4\pi} \left ( \sqrt{2}\operatorname{Im}\rho^2_{10}\sin2\vartheta\sin\varphi + \operatorname{Im}\rho^2_{1-1}\sin^2\vartheta\sin2\varphi \right )\nonumber
\end{eqnarray}

We construct the likelihood function
 \begin{eqnarray}\label{eq:likelihood}
        \ln L =  \sum_{i=1}^N \ln I(\Omega_i) - \sum_{j=1}^M \ln I(\Omega_j) - \int d\Omega\,I(\Omega)\,\eta(\Omega)
\end{eqnarray}

and use an extended-maximum likelihood fit to extract the SDMEs $\rho_{ij}^k$ such that the intensity fits the angular distribution of the observed $N$ events. The second term subtracts the contribution from accidental background that stems from combinations of events with unrelated photons. The third term is evaluated with a simulated sample that was generated without any angular dependence. This so-called normalization integral corrects for potential angular distortions by the detector acceptance. If the same Monte Carlo sample is weighted with the final SDMEs, it can be used to compare projections of angles between data and the fit results and therefore evaluate the fit results qualitatively.

\begin{figure}[t]
\centerline{\includegraphics[width=.9\textwidth]{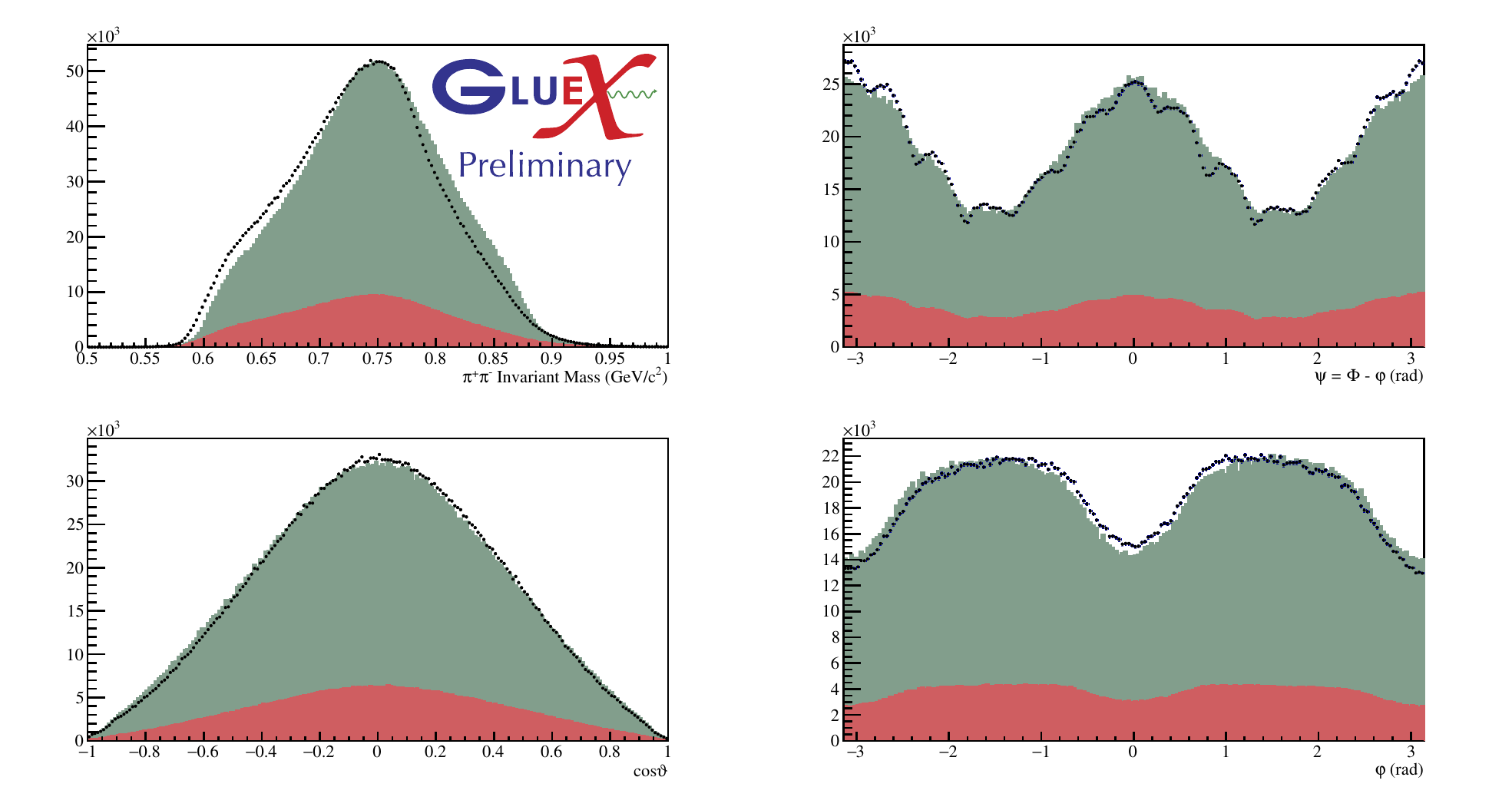}}
\caption{Evaluation of fit by comparison of measured distributions (black) to flat simulation weighted with fit results (green). The contribution from the subtracted accidental background is  shown in red. \label{fig:eval}}
\end{figure}

Figure~\ref{fig:eval} shows such a comparison for one orientation ($0^\circ$) and one example bin ($-t\in[0.05,0.15]\text{GeV}^2/c^2$).  The distributions in the angles $\cos\vartheta$, $\varphi$ and $\psi=\Phi-\varphi$ are very well reproduced. Small deviations indicate the contribution of background or shortcomings in the simulated model of the apparatus, which is still under active development.

\section{Results}
\subsection{$\rho(770) \rightarrow \pi^+\pi^-$}

\begin{figure}[ht]
\centerline{\includegraphics[width=\textwidth]{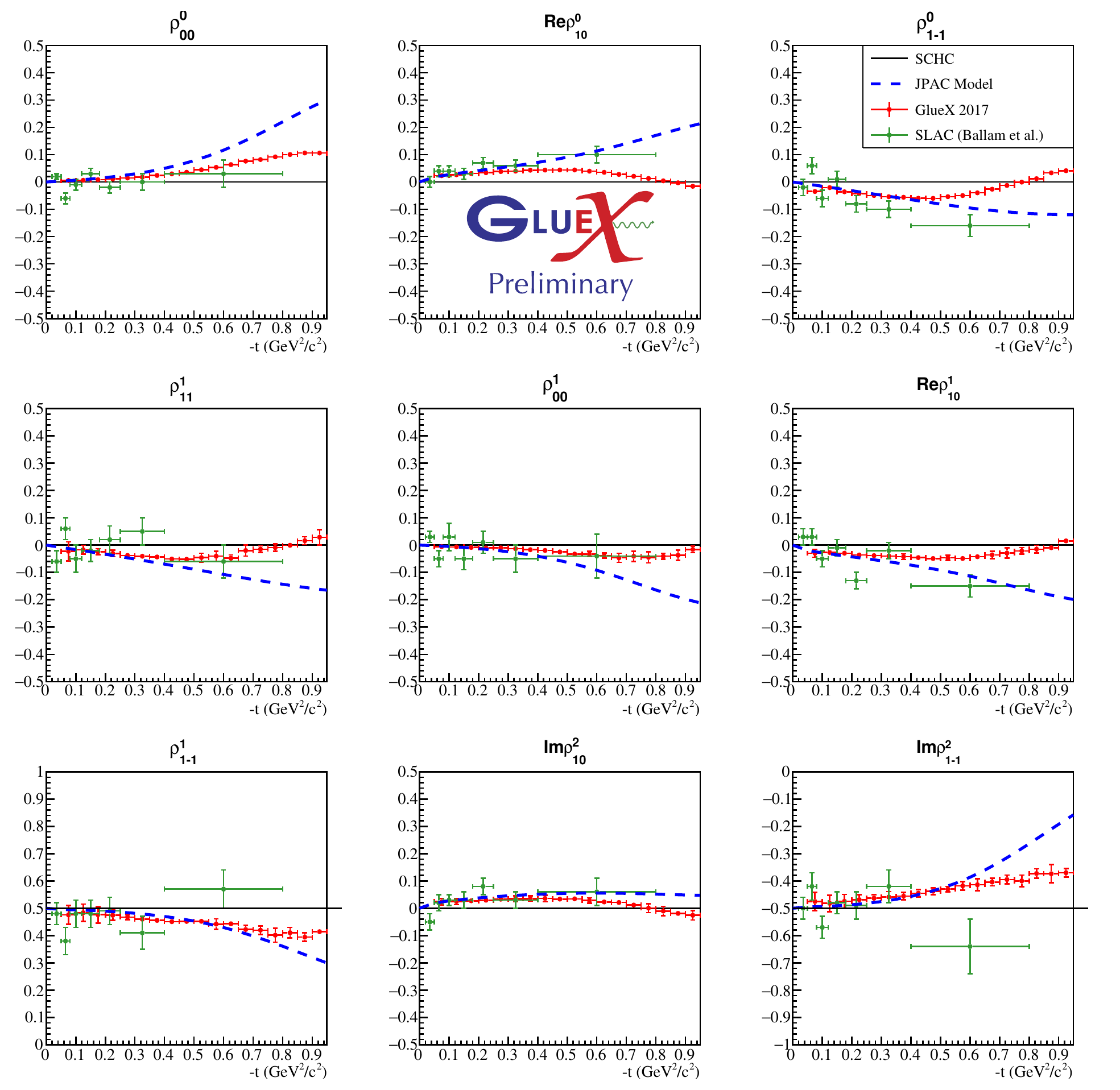}}
\caption{Spin-density matrix elements for the photoproduction of $\rho$(770) in the helicity frame.\label{fig:rhosdme}}
\end{figure}

For the $\rho$(770), we performed the analysis in 0.05\,GeV$^2/c^2$-wide bins in $-t$ separately for each orientation of the polarization direction. The result shown in Figure~\ref{fig:rhosdme} are the average of the 4 orientations, where the standard deviation from the mean value was used as a measure of the systematic uncertainties. Due to the magnitude of the data set, the statistical uncertainties are negligible.

In the limit of small $-t$, our results are consistent with the model of $s$-channel helicity conservation. Deviations from this model are predicted by Regge theory~\cite{mat18}, which our measurements closely follow up to around $-t\approx0.5\,$GeV$^2/c^2$. Above this point, the prediction loses its validity as an expansion in $\sqrt{m_0}/t$ is used, with $m_0=0.770\,$GeV$/c^2$ being the mass of the vector meson.

\begin{figure}[pt]
\centerline{\includegraphics[width=\textwidth]{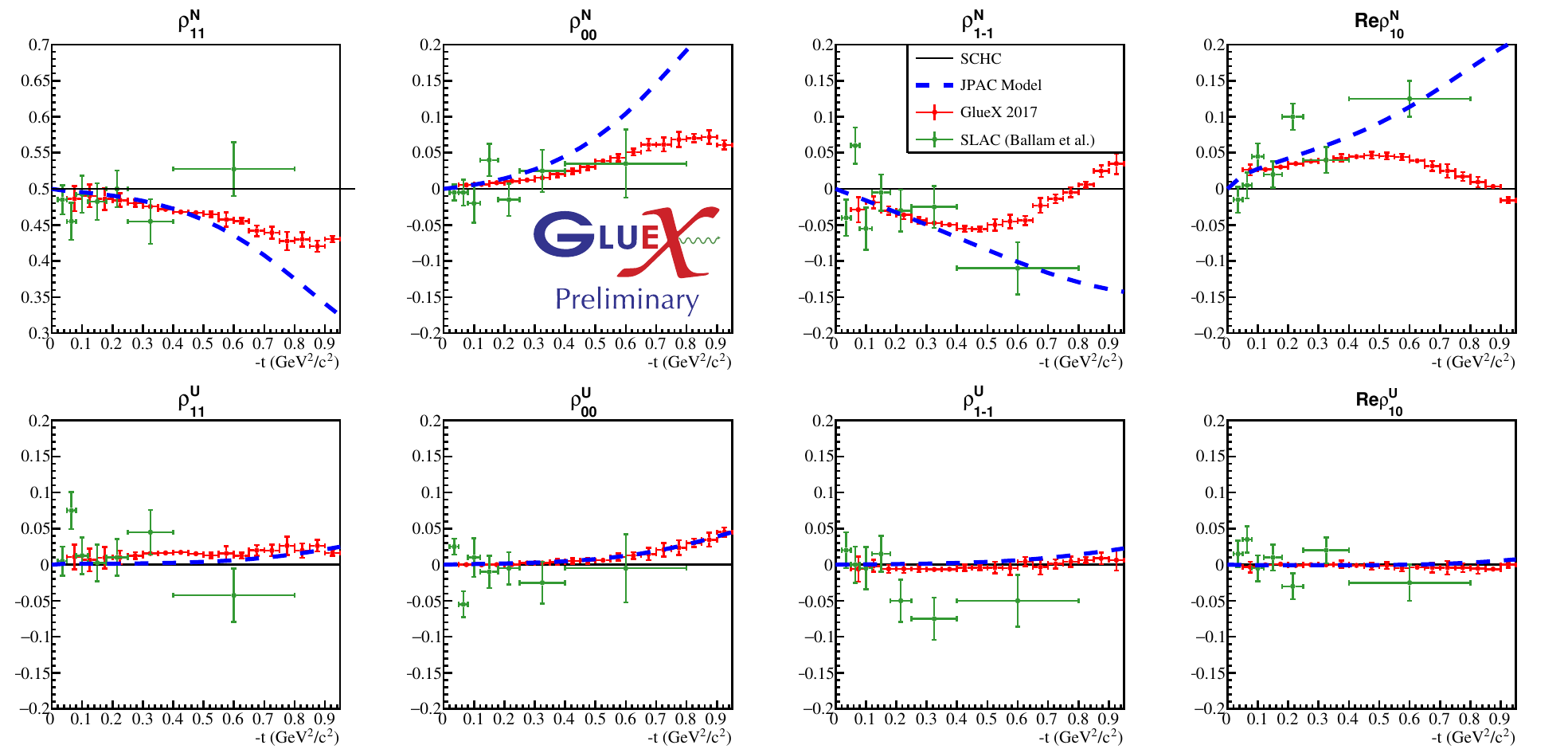}}
\caption{Spin-density matrix elements of natural- (top) and unnatural-parity exchange (bottom row) contributions.\label{fig:rhoparity}}
\end{figure}

The spin-density matrix can be separated into the components $\rho_{ik}^{\text{N,U}}$ arising from natural ($P = (-1)^J$) or unnatural ($P=-(-1)^J$) parity exchanges in the $t$ channel, respectively. The interference term between both production mechanisms vanishes in the limit of high energy~\cite{sch70}. We used the results from Figure~\ref{fig:rhosdme} to calculate the linear combinations
\begin{eqnarray}\label{eq:unnatural}
        \rho^{\text{N,U}}_{ik} = \tfrac{1}{2}(\rho_{ik}^0 \mp (-1)^i\rho_{-ik}^1)~.
\end{eqnarray}
Figure~\ref{fig:rhoparity} illustrates the clean separation. All unnatural components are compatible with zero, and the deviation from $s$-channel helicity conservation seem to originate from natural-parity exchange, which supports an earlier observation~\cite{bal73}.

In leading order, the asymmetry between natural and unnatural exchange cross sections can be reduced to one single observable, the parity asymmetry $P_\sigma$, which is defined as
\begin{eqnarray}\label{eq:psigma}
         P_\sigma = \frac{\sigma^N - \sigma^U}{\sigma^N + \sigma^U} = 2\rho^1_{1-1} - \rho^1_{00}.
\end{eqnarray}
In Figure~\ref{fig:psigma}a, we compare the calculated $P_\sigma$ with previous measurements and the Regge model. For $-t$ below $0.2\,\text{GeV}^2/c^2$, the results are consistent with unity, which confirms $s$-channel helicity conservation. The expected deviation from the Regge model for larger values of $-t$ is reflected in our results.

\subsection{$\phi(1020) \rightarrow K^+K^-$}

An analogous analysis has been performed for the $\phi$(1020) meson in the $K^+K^-$ final state. Data on this channel by previous measurements are very scarce, such that only a few hundred events were available to extract the SDMEs integrated over $-t$~\cite{bal73}. In contrast, only 20\% of the full GlueX data set allowed us to determine the spin-density matrix elements in 8 bins of 4-momentum transfer squared, which were chosen such that the number of events is roughly balanced. Figure~\ref{fig:psigma}b shows the calculated parity asymmetry. No significant $-t$ dependence is observed for the results. Agreement with the Regge model up to $-t\approx1\,\text{GeV}^2/c^2$ is expected due to the higher mass of the $\phi$(1020) meson.

\subsection{$\omega(782) \rightarrow \pi^+\pi^-\pi^0$}

\begin{figure}[ht]
    \begin{subfigure}{0.33\textwidth}
		\includegraphics[width=\textwidth]{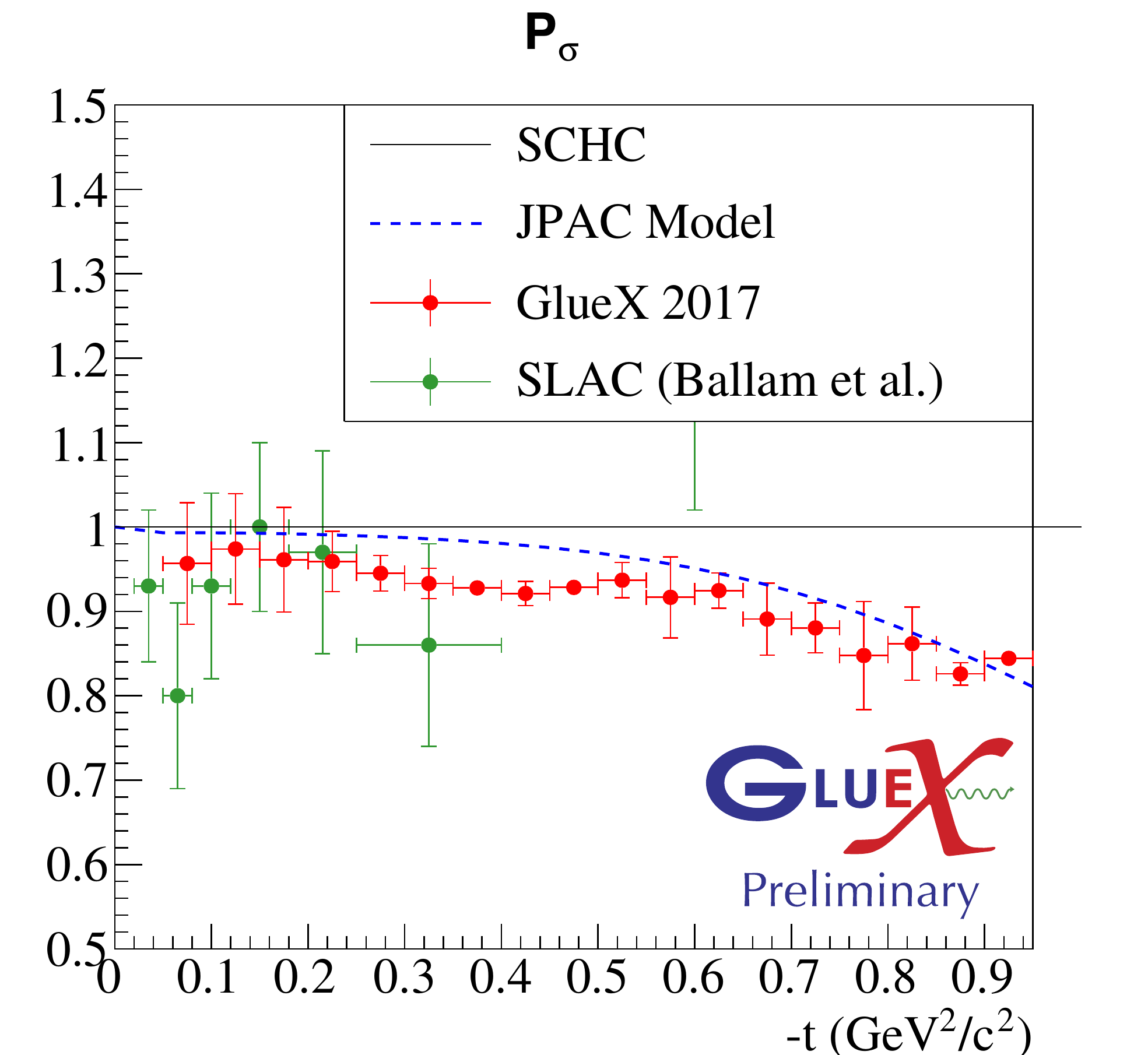}
		\caption{$\rho$(770)}
	\end{subfigure}
	\begin{subfigure}{0.33\textwidth}
		\includegraphics[width=\textwidth]{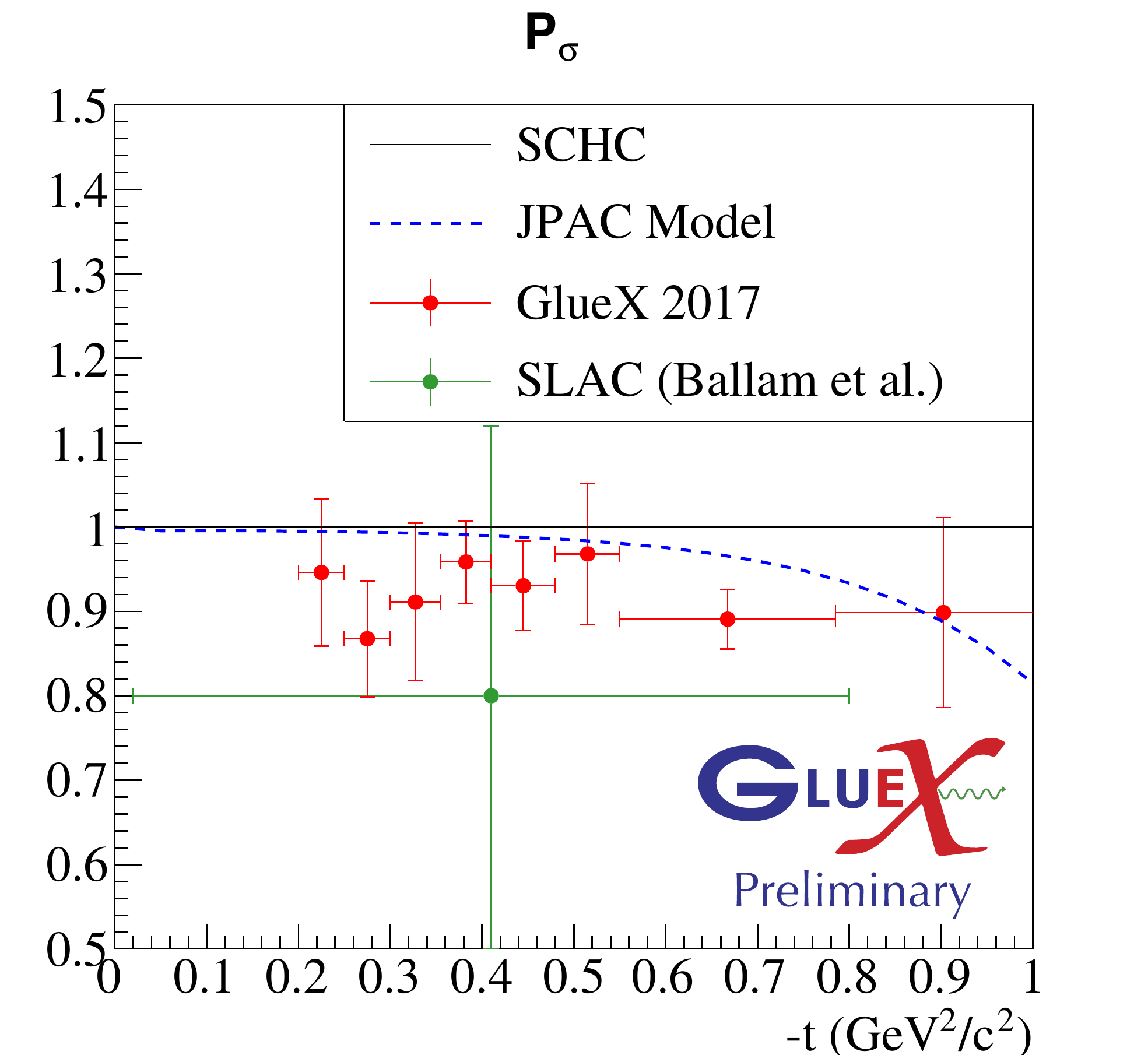}
		\caption{$\phi$(1020)}
	\end{subfigure}
	\begin{subfigure}{0.33\textwidth}
		\includegraphics[width=\textwidth]{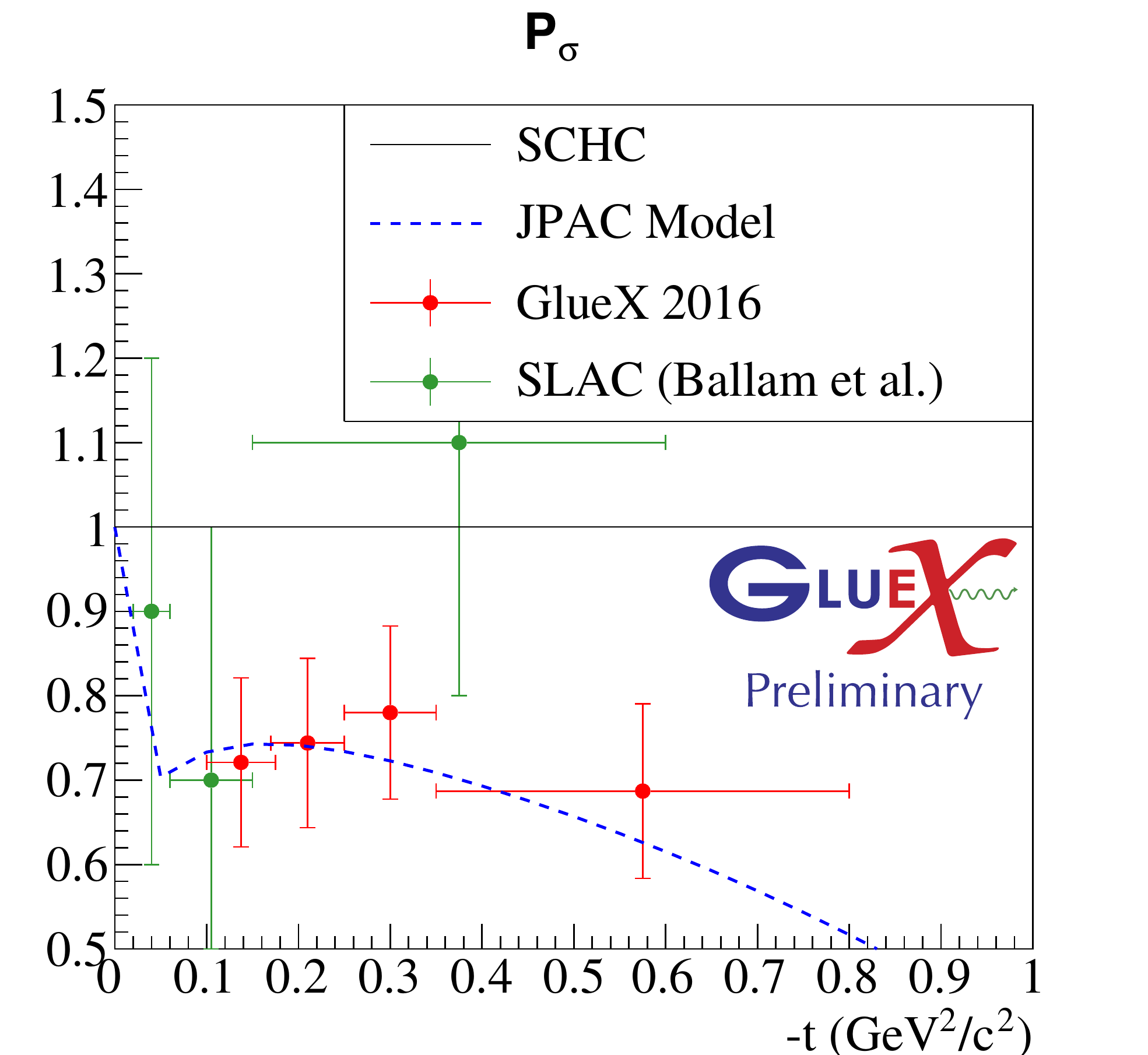}
		\caption{$\omega$(782)}
	\end{subfigure}
\caption{Parity asymmetry $P_\sigma$ for vector meson photoproduction.\label{fig:psigma}}
\end{figure}

We also analyze the decay of the $\omega$(782) meson into three pions in order to extract the spin-density matrix elements. Currently, only the results for a small sample of physics data that was recorded during a GlueX commissioning run in 2016 are available. The data sample was divided into 4 bins of $-t$ between 0.1 and 0.8$\,\text{GeV}^2/c^2$ with an approximately equal number of entries. The results shown in Figure~\ref{fig:psigma}c are the average of only two different orientations of the polarization plane. The parity asymmetry has a significant deviation from unity in this region. The significant deviation from unity with a value around 0.7 is consistent with the predictions from Regge theory~\cite{mat18}. Even with this limited data set, the statistical precision already exceeds that of previous measurements at comparable energies~\cite{bal73}.
We are currently extending the analysis to the full data set. The radiative decay of $\omega(782)$ to $\pi^0\gamma$ will be used as an independent confirmation.

\section{Summary and Outlook}

We report on the measurement of spin-density matrix elements of the vector mesons $\rho$(770), $\phi$(1020) and $\omega$(782) at GlueX. The statistical precision of the final analysis with the full data set will surpass previous measurements by orders of magnitude. In general, the production of the vector mesons with a $9\,\text{GeV}$ photon beam is consistent with $s$-channel helicity conservation only in the limit of $-t\rightarrow 0$. However, the decomposition of the spin-density matrix elements shows that natural parity exchanges dominate and the contribution from unnatural parity exchanges is small for the entire range in $-t$. This observation is consistent with predictions from Regge theory.

Results from this analysis serve as input to models of the production process, which will be essential for the interpretation of possible signals of exotic mesons in GlueX. In addition, these studies are used to study the GlueX detector and to evaluate its description in the simulation framework.

\begin{acknowledgments}

The Medium Energy Physics group at Carnegie Mellon University is supported by DOE Grant DE-FG02-87ER40315. The Thomas Jefferson National Accelerator Facility is supported by the U.S. Department of Energy, Office of Science, Office of Nuclear Physics under contract DE-AC05-06OR23177.


.

\end{acknowledgments}

\nocite{*}
\bibliography{aipsamp}

\providecommand{\noopsort}[1]{}\providecommand{\singleletter}[1]{#1}%
\begin{thebibliography}{7}%
\makeatletter
\providecommand \@ifxundefined [1]{%
 \@ifx{#1\undefined}
}%
\providecommand \@ifnum [1]{%
 \ifnum #1\expandafter \@firstoftwo
 \else \expandafter \@secondoftwo
 \fi
}%
\providecommand \@ifx [1]{%
 \ifx #1\expandafter \@firstoftwo
 \else \expandafter \@secondoftwo
 \fi
}%
\providecommand \natexlab [1]{#1}%
\providecommand \enquote  [1]{``#1''}%
\providecommand \bibnamefont  [1]{#1}%
\providecommand \bibfnamefont [1]{#1}%
\providecommand \citenamefont [1]{#1}%
\providecommand \href@noop [0]{\@secondoftwo}%
\providecommand \href [0]{\begingroup \@sanitize@url \@href}%
\providecommand \@href[1]{\@@startlink{#1}\@@href}%
\providecommand \@@href[1]{\endgroup#1\@@endlink}%
\providecommand \@sanitize@url [0]{\catcode `\\12\catcode `\$12\catcode
  `\&12\catcode `\#12\catcode `\^12\catcode `\_12\catcode `\%12\relax}%
\providecommand \@@startlink[1]{}%
\providecommand \@@endlink[0]{}%
\providecommand \url  [0]{\begingroup\@sanitize@url \@url }%
\providecommand \@url [1]{\endgroup\@href {#1}{\urlprefix }}%
\providecommand \urlprefix  [0]{URL }%
\providecommand \Eprint [0]{\href }%
\providecommand \doibase [0]{http://dx.doi.org/}%
\providecommand \selectlanguage [0]{\@gobble}%
\providecommand \bibinfo  [0]{\@secondoftwo}%
\providecommand \bibfield  [0]{\@secondoftwo}%
\providecommand \translation [1]{[#1]}%
\providecommand \BibitemOpen [0]{}%
\providecommand \bibitemStop [0]{}%
\providecommand \bibitemNoStop [0]{.\EOS\space}%
\providecommand \EOS [0]{\spacefactor3000\relax}%
\providecommand \BibitemShut  [1]{\csname bibitem#1\endcsname}%
\let\auto@bib@innerbib\@empty
\bibitem [{\citenamefont {Pooser}\ \emph {et~al.}(2019)\citenamefont {Pooser}
  \emph {et~al.}}]{poo19}%
  \BibitemOpen
  \bibfield  {author} {\bibinfo {author} {\bibfnamefont {E.}~\bibnamefont
  {Pooser}} \emph {et~al.},\ }\href@noop {} {\bibfield  {journal} {\bibinfo
  {journal} {Nucl.\ Instrum.\ and Meth. A}\ }\textbf {\bibinfo {volume}
  {927}},\ \bibinfo {pages} {330} (\bibinfo {year} {2019})}\BibitemShut
  {NoStop}%
\bibitem [{\citenamefont {Beattie}\ \emph {et~al.}(2018)\citenamefont {Beattie}
  \emph {et~al.}}]{Bea18}%
  \BibitemOpen
  \bibfield  {author} {\bibinfo {author} {\bibfnamefont {T.~D.}\ \bibnamefont
  {Beattie}} \emph {et~al.},\ }\href@noop {} {\bibfield  {journal} {\bibinfo
  {journal} {Nucl.\ Instrum.\ and Meth. A}\ }\textbf {\bibinfo {volume}
  {896}},\ \bibinfo {pages} {24} (\bibinfo {year} {2018})}\BibitemShut
  {NoStop}%
\bibitem [{\citenamefont {Dugger}\ \emph {et~al.}(2017)\citenamefont {Dugger}
  \emph {et~al.}}]{Dug17}%
  \BibitemOpen
  \bibfield  {author} {\bibinfo {author} {\bibfnamefont {M.}~\bibnamefont
  {Dugger}} \emph {et~al.},\ }\href@noop {} {\bibfield  {journal} {\bibinfo
  {journal} {Nucl.\ Instrum.\ and Meth. A}\ }\textbf {\bibinfo {volume}
  {867}},\ \bibinfo {pages} {115} (\bibinfo {year} {2017})}\BibitemShut
  {NoStop}%
\bibitem [{\citenamefont {Ghoul}\ \emph {et~al.}(2017)\citenamefont {Ghoul}
  \emph {et~al.}}]{Glx18}%
  \BibitemOpen
  \bibfield  {author} {\bibinfo {author} {\bibfnamefont {H.~A.}\ \bibnamefont
  {Ghoul}} \emph {et~al.},\ }\href@noop {} {\bibfield  {journal} {\bibinfo
  {journal} {Phys.\ Rev.\ C}\ }\textbf {\bibinfo {volume} {95}},\ \bibinfo
  {pages} {042201} (\bibinfo {year} {2017})}\BibitemShut {NoStop}%
\bibitem [{\citenamefont {Schilling}, \citenamefont {Seyboth},\ and\
  \citenamefont {Wolf}(1970)}]{sch70}%
  \BibitemOpen
  \bibfield  {author} {\bibinfo {author} {\bibfnamefont {K.}~\bibnamefont
  {Schilling}}, \bibinfo {author} {\bibfnamefont {P.}~\bibnamefont {Seyboth}},
  \ and\ \bibinfo {author} {\bibfnamefont {G.}~\bibnamefont {Wolf}},\
  }\href@noop {} {\bibfield  {journal} {\bibinfo  {journal} {Nucl.\ Phys.\ B}\
  }\textbf {\bibinfo {volume} {15}},\ \bibinfo {pages} {397} (\bibinfo {year}
  {1970})}\BibitemShut {NoStop}%
\bibitem [{\citenamefont {Ballam}\ \emph {et~al.}(1973)\citenamefont {Ballam}
  \emph {et~al.}}]{bal73}%
  \BibitemOpen
  \bibfield  {author} {\bibinfo {author} {\bibfnamefont {J.}~\bibnamefont
  {Ballam}} \emph {et~al.},\ }\href@noop {} {\bibfield  {journal} {\bibinfo
  {journal} {Phys.\ Rev.\ D}\ }\textbf {\bibinfo {volume} {7}},\ \bibinfo
  {pages} {3150} (\bibinfo {year} {1973})}\BibitemShut {NoStop}%
\bibitem [{\citenamefont {Mathieu}\ \emph {et~al.}(2018)\citenamefont {Mathieu}
  \emph {et~al.}}]{mat18}%
  \BibitemOpen
  \bibfield  {author} {\bibinfo {author} {\bibfnamefont {V.}~\bibnamefont
  {Mathieu}} \emph {et~al.},\ }\href@noop {} {\bibfield  {journal} {\bibinfo
  {journal} {Phys.\ Rev.\ D}\ }\textbf {\bibinfo {volume} {97}},\ \bibinfo
  {pages} {094003} (\bibinfo {year} {2018})}\BibitemShut {NoStop}%
\end{thebibliography}%

\end{document}